\documentclass[%
reprint,
showpacs,preprintnumbers,
 amsmath,amssymb,
 aps,
 pre,
twocolumn
]{revtex4-1}

\usepackage{graphicx}
\usepackage{dcolumn}
\usepackage{bm}
\usepackage{upgreek}
\usepackage{xcolor}

\begin{document}

\preprint{APS/123-QED}

\title{High-resolution Monte Carlo study of the order-parameter distribution of the three-dimensional Ising model}

\author{Jiahao Xu$^1$}
\email[]{jiahaoxu@uga.edu}
\author{Alan M. Ferrenberg$^2$}
\email[]{alan.ferrenberg@miamioh.edu}
\author{David P. Landau$^1$}
\email[]{dlandau@physast.uga.edu}
\affiliation{$^1$Center for Simulational Physics, University of Georgia, Athens, GA 30602 USA\\
$^2$Department of Computer Science and Software Engineering, Miami University of Ohio, Oxford, OH 45056 USA}

\date{\today}

\begin{abstract}
We apply extensive Monte Carlo simulations to study the probability
distribution $P(m)$ of the order parameter $m$ for the simple cubic
Ising model with periodic boundary condition at the transition
point. Sampling is performed with the Wolff cluster flipping
algorithm, and histogram reweighting together with finite-size scaling analyses are then used to extract a precise functional form for the probability distribution of the magnetization, $P(m)$, in
the thermodynamic limit.  This form should serve as a benchmark for other models in the three-dimensional Ising Universality class. 


\end{abstract}



\maketitle


\section{\label{sec:level1}introduction}

The probability distribution $P(m)$ of the order parameter $m$ is one
of the most important quantities for studying the finite-size scaling
of critical phenomena.  It contains the information needed to
calculate all order parameter related quantities such as the susceptibility $\chi
= K ( \left< m^2 \right> - \left< m \right>^2 )$, where $K$ is the
dimensionless inverse temperature, the Binder cumulant $U = 1 - \left<
m^4 \right> / (3 \left< m^2 \right>^2)$, etc.  It can also complement
the use of critical exponents in determining the critical behavior of
a universality class.  For these reasons it has been a major research topic
in multiple Monte Carlo studies~\cite{Binder1981, Bruce1992, Blote1995, Kim1996, cross_correlation2010}. With precise calculations of these
quantities, one can study the transition temperature and critical behavior
of diverse systems, e.g.
the 3D Ising Model~\cite{Ferrenberg1991, 3dIsing2018}, 
the Lennard-Jones fluid~\cite{Wilding1995}, and
quantum chromodynamics~\cite{QCD2010}. A very nice application of the magnetization probability distribution function to determine the critical and multicritical universality in several different spin systems can be found in Ref.~\cite{PLASCAK2013259}.


According to finite-size scaling theory~\cite{Binder1981,Hilfer1995}, and assuming hyperscaling and using $L$ (linear dimension), $m$ (order parameter), and $\xi$ (correlation length) as variables, the probability distribution of the order parameter is described by the scaling ansatz,
\begin{equation}
P(m, L, \xi) = L^{\beta / \nu} \tilde{P}(m L^{\beta / \nu}, L / \xi)
\end{equation}
where $\beta$ is the order parameter exponent, $\nu$ is the correlation length exponent, and $\tilde{P}(m L^{\beta / \nu}, L / \xi)$ is the scaling function.

The double peaked distribution of $P(m)$ for the simple cubic Ising
model was first numerically calculated by Monte Carlo simulation in
Ref.~\cite{Binder1981}. In Ref.~\cite{Hilfer1995}, systems of size
$20^3$ and $30^3$ were simulated at the critical point and an
analytical expression for $P(m)$ was proposed.  An improved
estimate for $P(m)$ was determined in Ref.~\cite{pm_Tsypin}, where the
size of the simple cubic lattices ranged from $12^3$ to $58^3$. That
work established a phenomenological formula to describe the peaks of the
distribution. In addition to the Ising model, this study tried to extract $P(m)$ in the thermodynamic limit from simulations of the simple cubic, spin-1 Blume-Capel model.
The tail of the  probability distribution $P(m)$ for the $2D$
Ising model was studied in Ref.~\cite{Hilfer2003}, but the conclusion was
that the true form of the order parameter distribution at criticality
was still an open question. 

High-resolution numerical estimates for properties of $P(m)$ are
important for developing theories and analytical methods for the study of critical phenomena. Our goal in
the present paper is to determine the probability distribution of the
order parameter at the critical point of the simple cubic Ising model with increased resolution and
obtain a more precise expression to describe $P(m)$ in the
thermodynamic limit than was heretofore possible.

\section{\label{sec:level2}model and methods}

We consider the $3D$ Ising model on a simple cubic lattice with linear dimension $L$ and periodic boundary conditions. The Hamiltonian is given
by,
\begin{equation}
\mathcal{H} = -J\sum_{\langle i,j \rangle}\sigma_i\sigma_j, \;\;\;\; \sigma_i = \pm 1
\end{equation}
Here $J > 0$ is the ferromagnetic coupling, $\langle i,j \rangle$
denotes pairs of nearest-neighbor sites, and the sum is over
the $3N$ distinct pairs of nearest-neighbors, where $N = L^3$ is the
total number of spins. The order parameter (average magnetization) is given by
\begin{equation}
m = \frac{1}{N} \sum_{i} \sigma_i
\end{equation}
where $i$ denotes each of the $N$ spins, and $-1 \leq m \leq 1$.

We performed extensive Monte Carlo simulations using the Wolff cluster flipping
algorithm~\cite{Wolff}.
The simulations were performed at $K_0=0.221\,654$, which was an
estimate for the inverse critical temperature used in an
earlier, high resolution Monte Carlo study~\cite{Ferrenberg1991}. Data
were obtained for lattices with $16 \leq L \leq 1024$ (for more
simulation details, see Ref.~\cite{3dIsing2018}). 

%

Based on the estimate for the critical point in
Ref.~\cite{3dIsing2018}, data were reweighted to $K_c =
0.221\,654\,626$ using histogram reweighting
techniques~\cite{histogram_Ferrenberg1988,
  histogram_Ferrenberg1989}. To obtain the probability distribution
$P(m)$ at $K_c$, for each occurrence of the order parameter, the
corresponding population of the bin of the histogram was incremented by
$\exp(-(K_c - K_0)E)$, where $E$ is the total dimensionless energy of
the system. The histogram was then normalized to determine $P(m)$.

\section{\label{sec:level3}results}
\begin{figure}
\centering
\includegraphics [width=0.95\hsize] {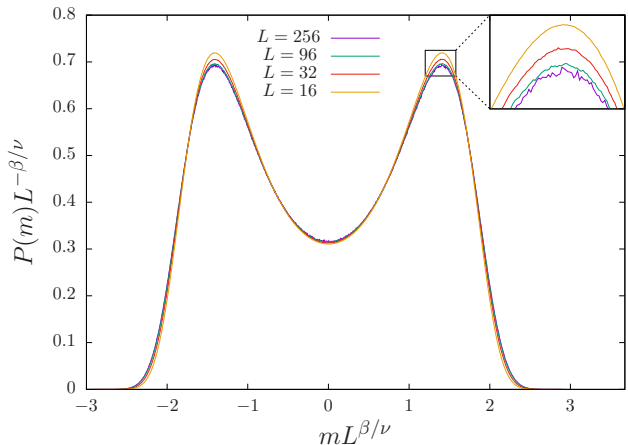}
\caption{(color online) Scaled probability distribution $P(m) L^{-\beta / \nu}$ as a function of $m L^{\beta / \nu}$ at the critical point $K_c = 0.221\,654\,626$~\cite{3dIsing2018}. Curves from the top to the bottom in the inset correspond to lattice sizes $L = 16$, $32$, $96$, and $256$ respectively.}
\label{fig_pm}
\end{figure}

Fig.~\ref{fig_pm} shows the scaled probability distribution $P(m)L^{-\beta / \nu}$ as a function of $m L^{\beta / \nu}$ at the critical
point $K_c = 0.221\,654\,626$ for finite lattice sizes ($L = 16$,
$32$, $96$, and $256$). Here, $\beta$ and $\nu$ are critical exponents
for infinite lattices, and $\beta / \nu =
0.518\,01(35)$~\citep{3dIsing2018}. (We used this estimate in our analysis for consistency since this work uses the same data as that of Ref.~\cite{3dIsing2018}.) The values of the scaled peaks
$P(m) L^{-\beta / \nu}$ decreases as the lattice size $L$ 
increases. Also, systematic deviations from scaling occur in the region of the tails of the distributions.
In the thermodynamic limit ($L = \infty$), the probability
distribution $P(m)$ is universal up to a rescaling of $m$.

First, we took Ref.~\cite{pm_Tsypin} as the blueprint for our analysis.
We performed a nonlinear least-squares fit, where the
reciprocals of the statistical errors were taken as the weighting
factors to the fitting function, with the ``improved'' ansatz of
Ref.~\cite{pm_Tsypin},
\begin{align}
\label{ansatz1}
&P(m) = A L^{\beta / \nu} \nonumber \\
&\times \exp \left\{ - \left[ \left( \frac{m L^{\beta / \nu}} {m_0} \right)^2 - 1 \right]^2 \left[ b \left( \frac{m L^{\beta / \nu}} {m_0} \right)^2 + c \right] \right\}
\end{align}
where $A$, $m_0$, $b$, and $c$ are unknown fitting parameters. Note
that $m_0$ is a scale-invariant (but not universal) quantity.

\begin{figure}
\centering
\includegraphics [width=0.95\hsize] {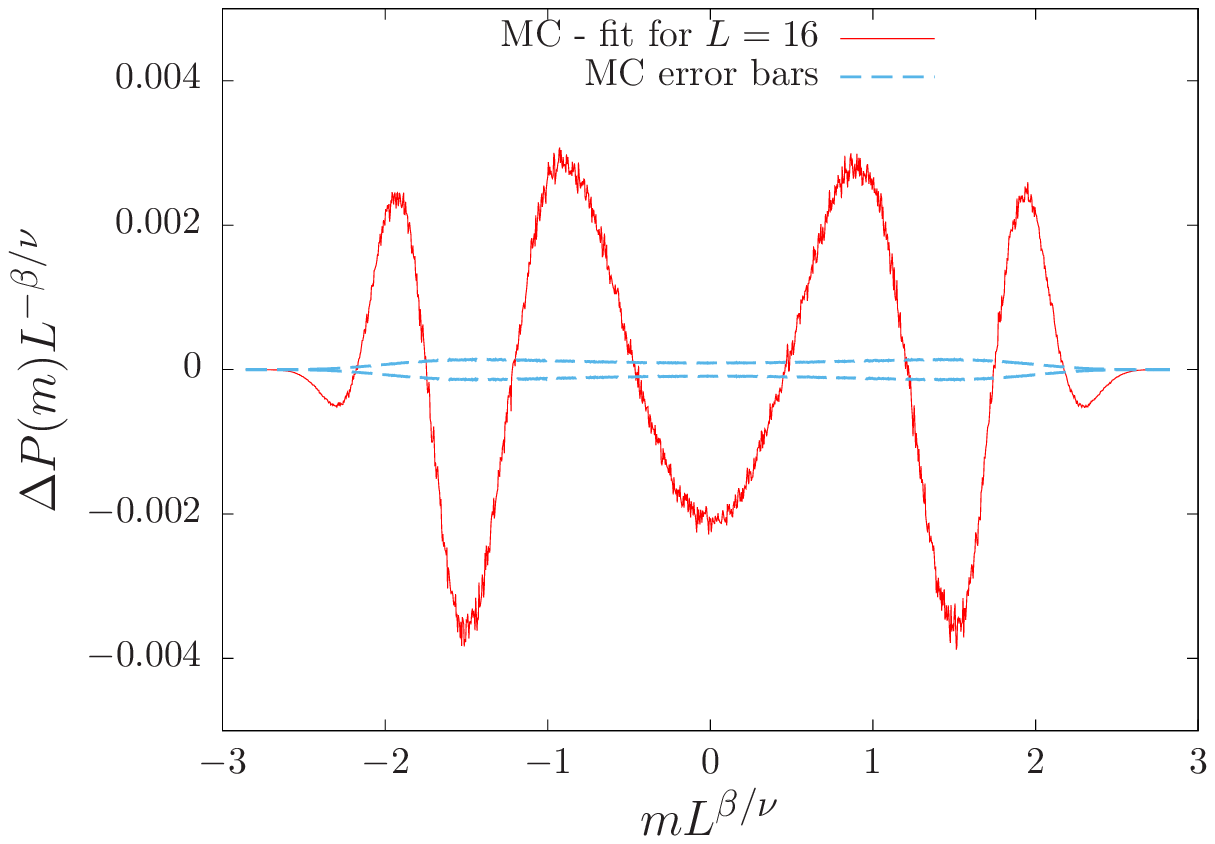}
\includegraphics [width=0.95\hsize] {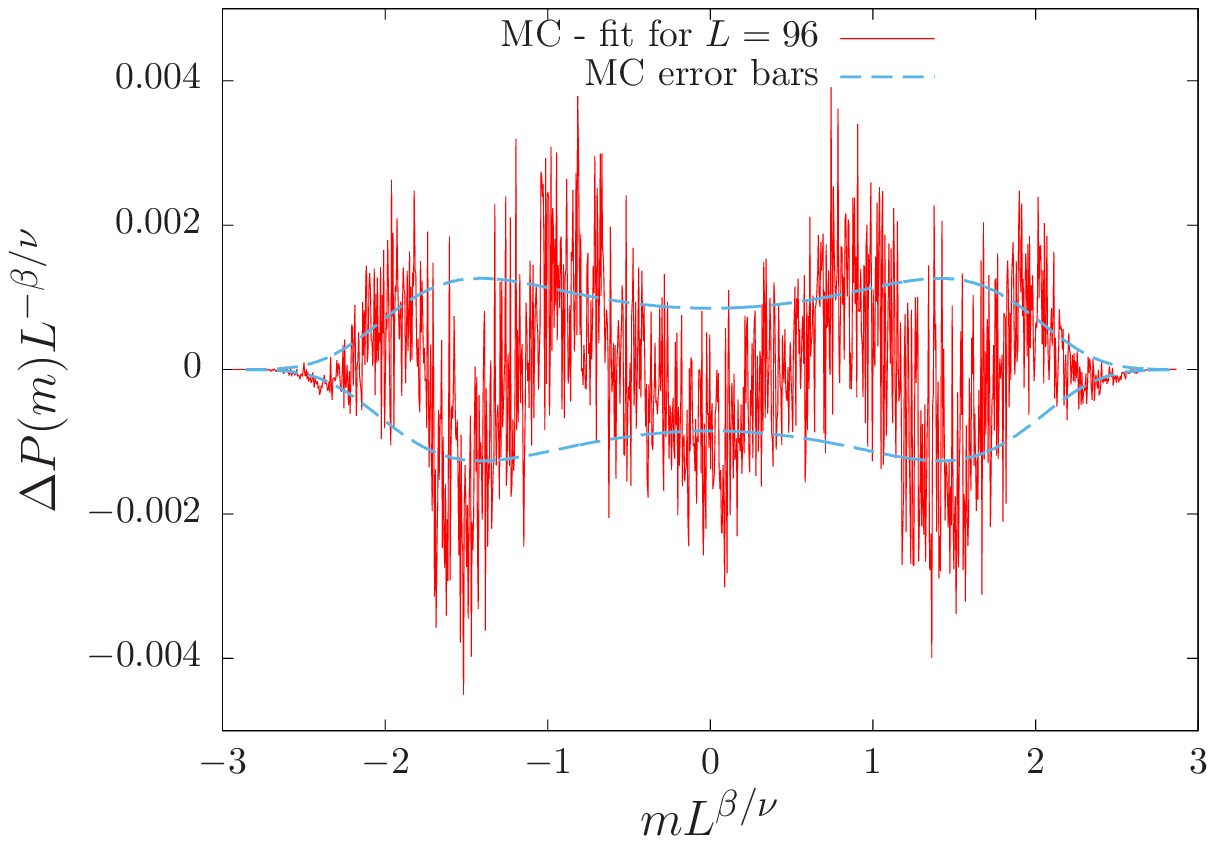}
\includegraphics [width=0.95\hsize] {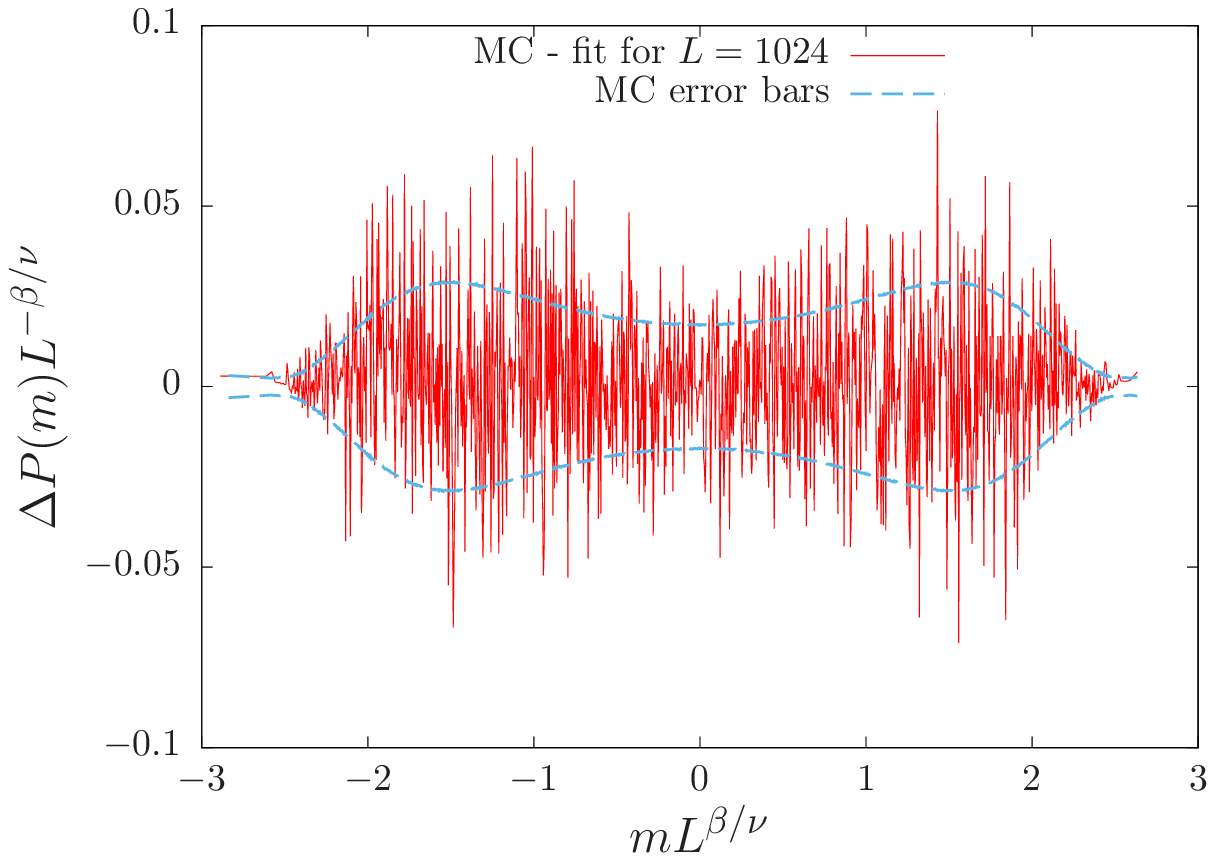}
\caption{(color online) The red (dark grey) line is the difference between the Monte Carlo data and the fit corresponding to Eq.~(\ref{ansatz1}), while the blue (light grey) line is the error bar for the Monte Carlo data (top: $L = 16$, middle: $L = 96$, bottom: $L = 1024$). }
\label{fig_ansatz1}
\end{figure}

Fig.~\ref{fig_ansatz1} shows the difference between the Monte Carlo(MC)
data and the fit corresponding to Eq.~(\ref{ansatz1}). It also
illustrates the error bars for the Monte Carlo data.  From
Fig.~\ref{fig_ansatz1} we observe that when the lattice size $L$ is
small, e.g. $L = 16$, a pattern in the difference between MC data
and the fit is very clear. This means that the fitting ansatz,
Eq.~(\ref{ansatz1}), does not perform well for small $L$ to within the
statistical uncertainty. For much larger $L$ (e.g. $L = 1024$), the
difference between the distribution and the fit to the ansatz is of
the same magnitude as the statistical error, so no systematic deviation
is observed.

Table~\ref{table_ansatz1} shows the results of fitting to
Eq.~(\ref{ansatz1}). We can tell that the quality of fit is not good
when $L \leq 80$, as the value of the $\chi^2$ per degree of freedom
(d.o.f.) is large. It decreases for larger $L$, and the quality of
fit becomes good for the largest lattice sizes. 

\begin{table*} 
\caption{The parameters $m_0$, $b$, and $c$ for the probability
  distribution $P(m)$, fitted to the ansatz Eq.~(\ref{ansatz1}). The
  last column $\chi^2$ per degree of freedom (d.o.f.) characterizes
  the quality of the fit.}
\begin{ruledtabular}
\begin{tabular}{@{\hspace{7em}}c c c c c@{\hspace{7em}}}
$L$ & $m_0$ & $b$ & $c$ & $\chi^2$ per d.o.f. \\ \hline
16	& 1.411 97 (26)		& 0.2408 (17)  	& 0.836 86 (77) &	381.82 \\
24	& 1.411 12 (12)		& 0.209 24 (70)	& 0.819 38 (40) &	70.96 \\
32	& 1.410 761 (82)	& 0.195 20 (48)	& 0.810 45 (21) &	24.87 \\
48	& 1.410 437 (74)	& 0.181 97 (43)	& 0.800 87 (25) &	6.98 \\
64	& 1.410 440 (46)	& 0.176 01 (29)	& 0.796 07 (26) &	3.70 \\
80	& 1.410 351 (48)	& 0.172 20 (31)	& 0.793 03 (33) &	2.23 \\
96	& 1.410 345 (57)	& 0.169 77 (35)	& 0.791 04 (34) &	1.74 \\
112	& 1.410 250 (59)	& 0.167 85 (37)	& 0.789 39 (42) &	1.47 \\
128	& 1.410 362 (71)	& 0.166 74 (46)	& 0.788 09 (36) &	1.32 \\
144	& 1.410 153 (85)	& 0.165 37 (54)	& 0.786 93 (37) &	1.24 \\
160	& 1.410 217 (98)	& 0.164 62 (62)	& 0.786 39 (42) &	1.18 \\
192	& 1.410 189 (67)	& 0.163 36 (85)	& 0.784 99 (46) &	1.12 \\
256	& 1.410 281 (87)	& 0.1620 (11)	& 0.783 59 (47) &	1.08 \\ 
384	& 1.410 18 (11)		& 0.1560 (14)	& 0.781 98 (49) &	1.04 \\
512	& 1.410 19 (21)		& 0.1590 (18)	& 0.781 01 (63) &	1.02 \\ 
768	& 1.410 97 (56)		& 0.1576 (47)	& 0.781 65 (70) &	1.02 \\
1024& 1.410 84 (75)		& 0.1545 (89)	& 0.780 76 (84) &	1.01 \\

\end{tabular}
\end{ruledtabular}
\label{table_ansatz1}
\end{table*}

\begin{table*} 
\caption{The parameters $m_0$, $a$, $b$, and $c$ for the probability
  distribution $P(m)$, fitted by the ansatz Eq.~(\ref{ansatz2}). The
  last column $\chi^2$ per degree of freedom (d.o.f.) characterizes
  the quality of the fit.}
\begin{ruledtabular}
\begin{tabular}{@{\hspace{2em}}c c c c c c@{\hspace{2em}}}
$L$ & $m_0$ & $a$ & $b$ & $c$ & $\chi^2$ per d.o.f. \\ \hline
16 &	1.408 684 (19)&	0.025 01 (21) &	0.169 36 (31) &	0.839 36 (33) &	1.31 \\
24 &	1.408 497 (27)&	0.016 44 (15) &	0.160 64 (23) &	0.821 24 (36) &	1.03 \\
32 &	1.408 456 (44)&	0.013 10 (14) &	0.155 53 (12) &	0.812 03 (26) &	1.04 \\
48 &	1.408 432 (61)&	0.010 46 (15) &	0.150 02 (24) &	0.802 20 (27) &	1.02 \\
64 &	1.408 588 (49)&	0.009 26 (21) &	0.147 51 (29) &	0.797 27 (31) &	1.03 \\
80 &	1.408 573 (53)&	0.008 62 (25) &	0.145 55 (40) &	0.794 16 (39) &	1.02 \\
96 &	1.408 611 (73)&	0.008 22 (27) &	0.144 27 (54) &	0.792 13 (45) &	1.02 \\
112 &	1.408 564 (70)&	0.007 84 (28) &	0.143 47 (75) &	0.790 43 (51) &	1.01 \\
128 &	1.408 714 (64)&	0.007 54 (36) &	0.143 26 (61) &	0.789 10 (48) &	1.02 \\
144 &	1.408 490 (82)&	0.007 48 (45) &	0.142 07 (76) &	0.787 93 (55) &	1.02 \\
160 &	1.408 580 (92)&	0.007 27 (56) &	0.141 94 (93) &	0.787 36 (48) &	1.01 \\
192 &	1.408 497 (91)&	0.007 31 (52) &	0.140 53 (99) &	0.785 97 (42) &	1.01 \\
256 &	1.408 672 (95)&	0.006 75 (65) &	0.1409 (12) &	0.784 48 (58) &	1.01 \\
384 &	1.408 489 (87)&	0.006 57 (74) &	0.1396 (20) &	0.782 80 (89) &	1.01 \\
512 &   1.408 52 (12) & 0.006 35 (93) & 0.1395 (25) &   0.7817 (13) & 1.01 \\
768 &   1.408 82 (19) & 0.0052 (17)   & 0.1423 (61) &   0.7821 (17) & 1.01 \\
1024&   1.408 73 (27) & 0.0043 (27)   & 0.1440 (90) &   0.7806 (25) & 1.01 \\

\end{tabular}
\end{ruledtabular}
\label{table_ansatz2}
\end{table*}

\begin{figure}
\centering
\includegraphics [width=0.95\hsize] {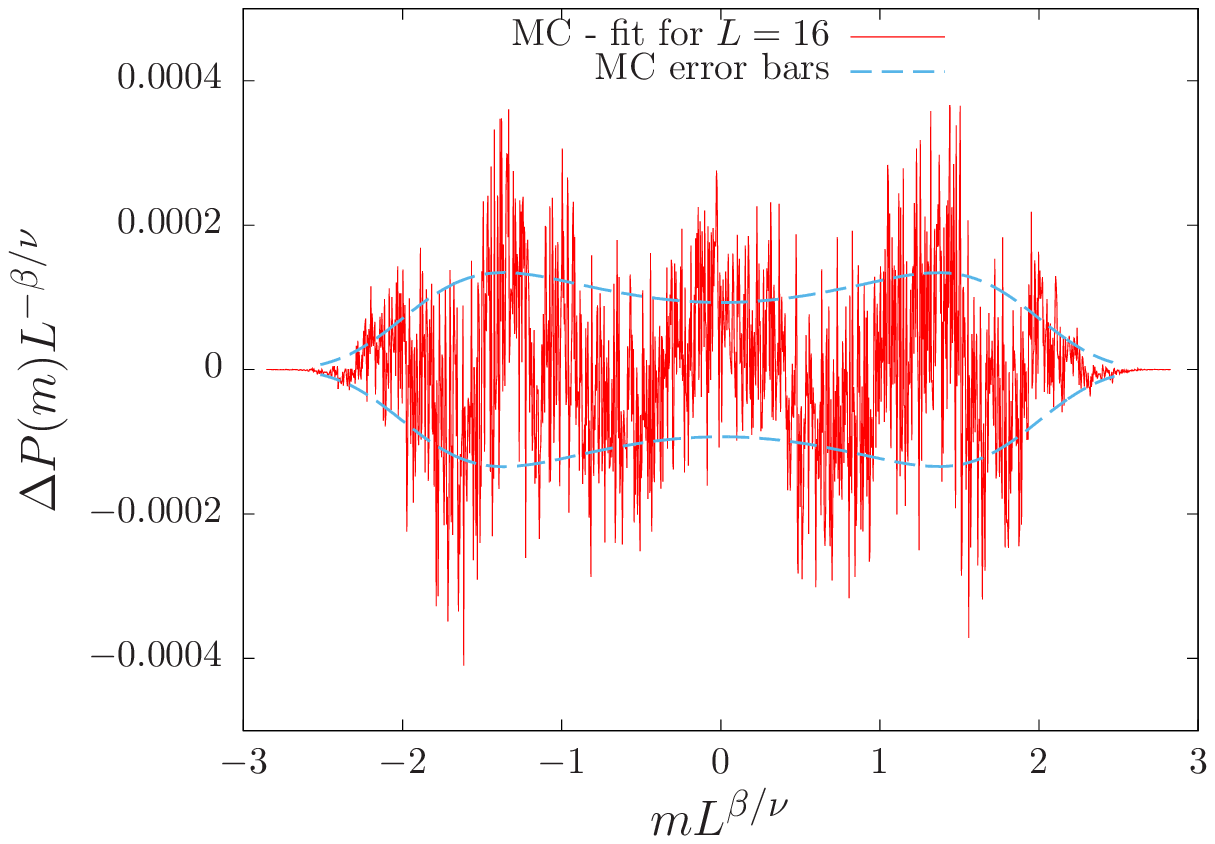}
\includegraphics [width=0.95\hsize] {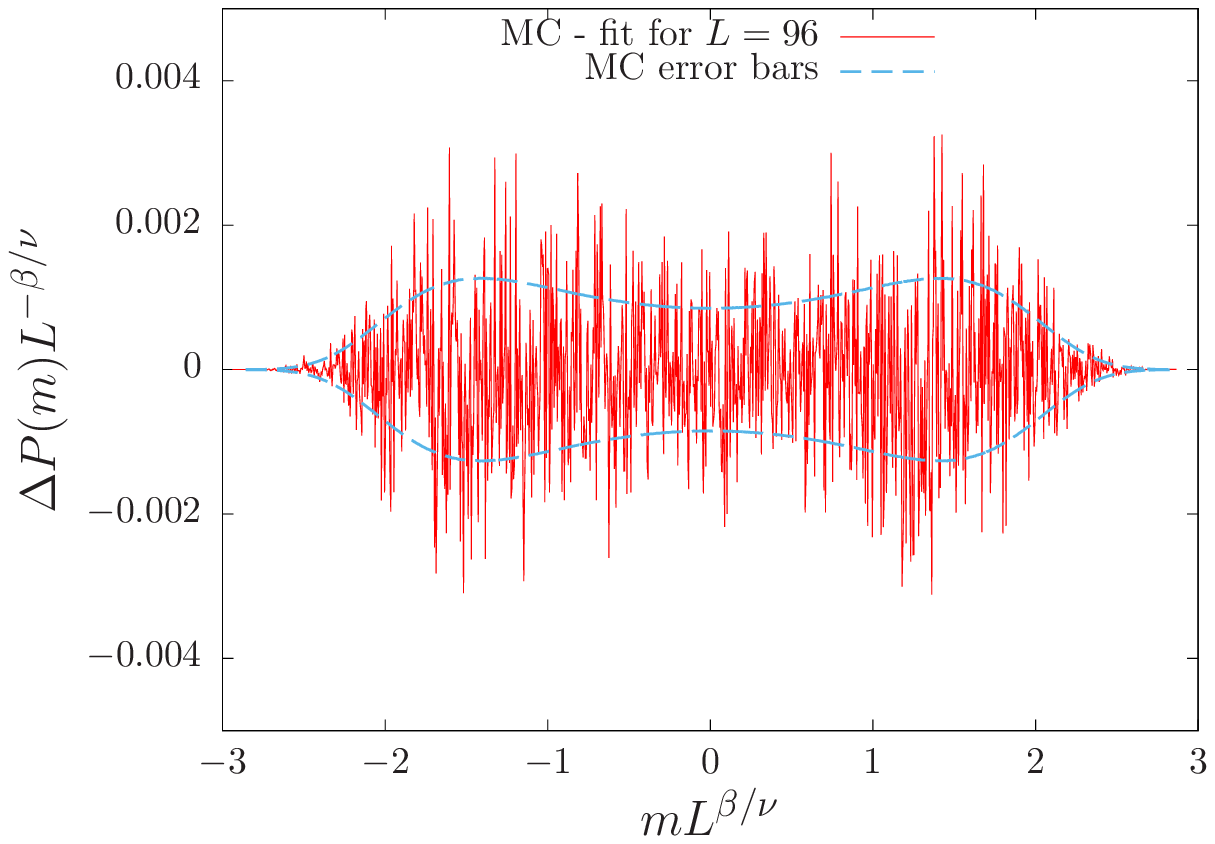}
\includegraphics [width=0.95\hsize] {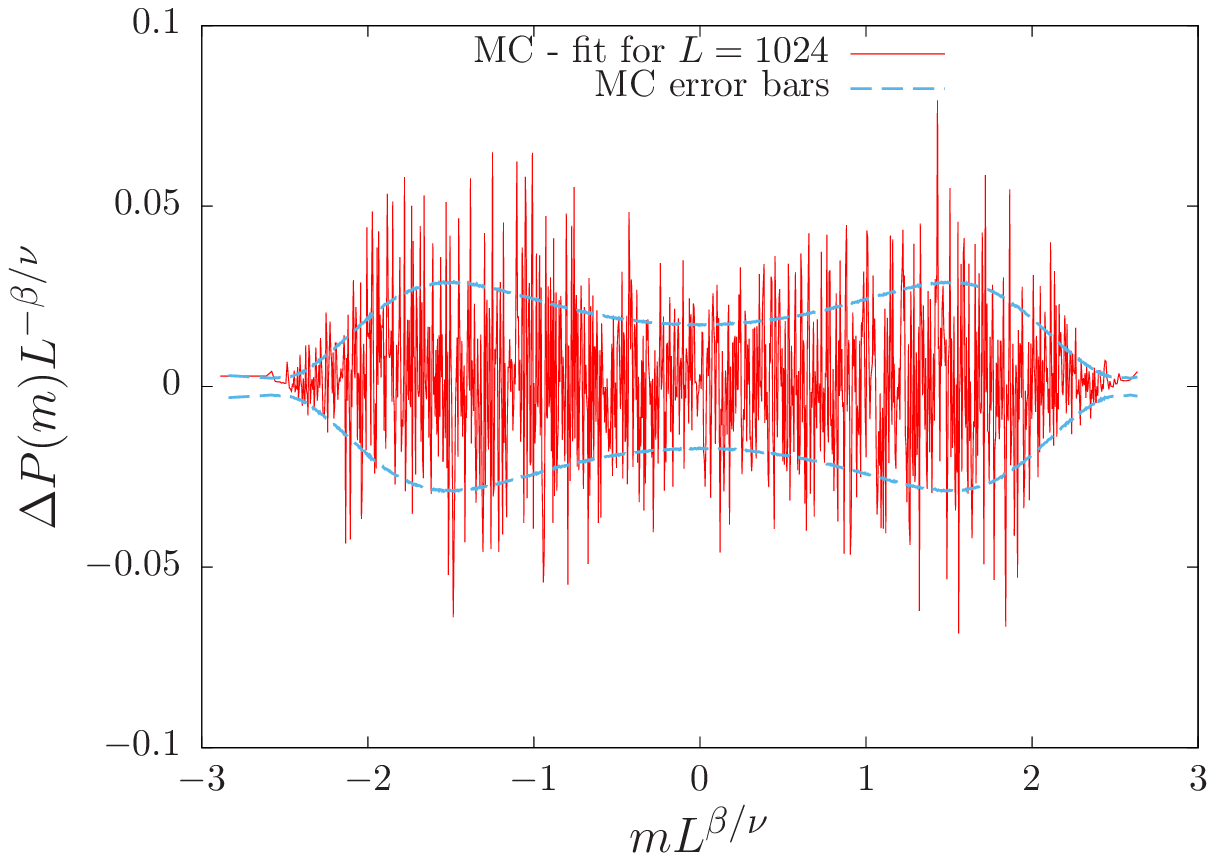}
\caption{(color online) Analogous to Fig.~\ref{fig_ansatz1}, but the fit is corresponding to Eq.~(\ref{ansatz2}) (top: $L = 16$, middle: $L = 96$, bottom: $L = 1024$). }
\label{fig_ansatz2}
\end{figure}

Based on the variance of the fit parameters $b$ and $c$ of
Eq.~(\ref{ansatz1}) for different lattice sizes, we have estimated
their values and errors for $L = \infty$ as follows, 
\begin{equation}
\label{result1}
b = 0.1553(6), \;\; c = 0.7783(4)
\end{equation}
Ref.~\cite{pm_Tsypin} determined the less precise values $b = 0.158(2)$ and $c = 0.776(2)$ which agree with our results within the error bars.

The systematic deviation observed for smaller system sizes led us to
modify ansatz Eq.~(\ref{ansatz1}) by adding various forms of
correction terms to see if a revised ansatz could fit the data well
even for smaller lattices. We approximated $P(m)$ by using
different forms, e.g. adding correction terms in the exponent,
adding different correction terms in the pre-exponential factor
($|m|^{\omega}, |m|, |m|^2, ...$), and adding correction terms in both
the exponent and the pre-exponent factor. We have found that the
following ``improved'' ansatz gives a surprisingly good approximation to $P(m)$ over quite a wide range of $L$ and $m$:
\begin{widetext}
\begin{equation}
\label{ansatz2}
P(m) = A L^{\beta / \nu} \exp \left\{ - \left[ \left( \frac{m L^{\beta / \nu}} {m_0} \right)^2 - 1 \right]^2 \left[ a \left( \frac{m L^{\beta / \nu}} {m_0} \right)^4 + b \left( \frac{m L^{\beta / \nu}} {m_0} \right)^2 + c \right] \right\}
\end{equation}
\end{widetext}
where $A$, $m_0$, $a$, $b$, and $c$ are unknown fit parameters,
and as before $\beta / \nu = 0.51801(35)$.

Fig.~\ref{fig_ansatz2} is analogous to Fig.~\ref{fig_ansatz1}, but
shows the difference between the Monte Carlo data and the fits to
Eq.~(\ref{ansatz2}). Fig.~\ref{fig_ansatz2} shows that
even for $L = 16$ the residual discrepancy is comparable to the
statistical error. If Eq.~(\ref{ansatz2}) is used as the fitting
function, the maximal difference between MC data and the fit for $L =
16$ is around $0.0004$, which is $1/10$ of that in
Fig.\ref{fig_ansatz1} which used Eq.~(\ref{ansatz1}) as the fitting
function. Thus, the quality of fitting to ansatz Eq.~(\ref{ansatz2})
is much higher than that of Eq.~(\ref{ansatz1}) for small $L$, and 
within the statistical errors, Eq.~(\ref{ansatz2}) performs better
than Eq.~(\ref{ansatz1}) as a fitting function.

Results for fitting to the functional form Eq.~(\ref{ansatz2}) are
shown in Table~\ref{table_ansatz2}. The values of the $\chi^2$ per
d.o.f. show that the quality of fit is good even for small
lattice sizes. Generally speaking, the error bars for the fit
parameters ($m_0$, $a$, $b$, and $c$) become larger as $L$
increases. This is because the statistical errors of the raw data are
greater for larger lattice sizes (see the dashed line in
Fig.~\ref{fig_ansatz2}). 

\begin{figure}
\centering
\includegraphics [width=0.95\hsize] {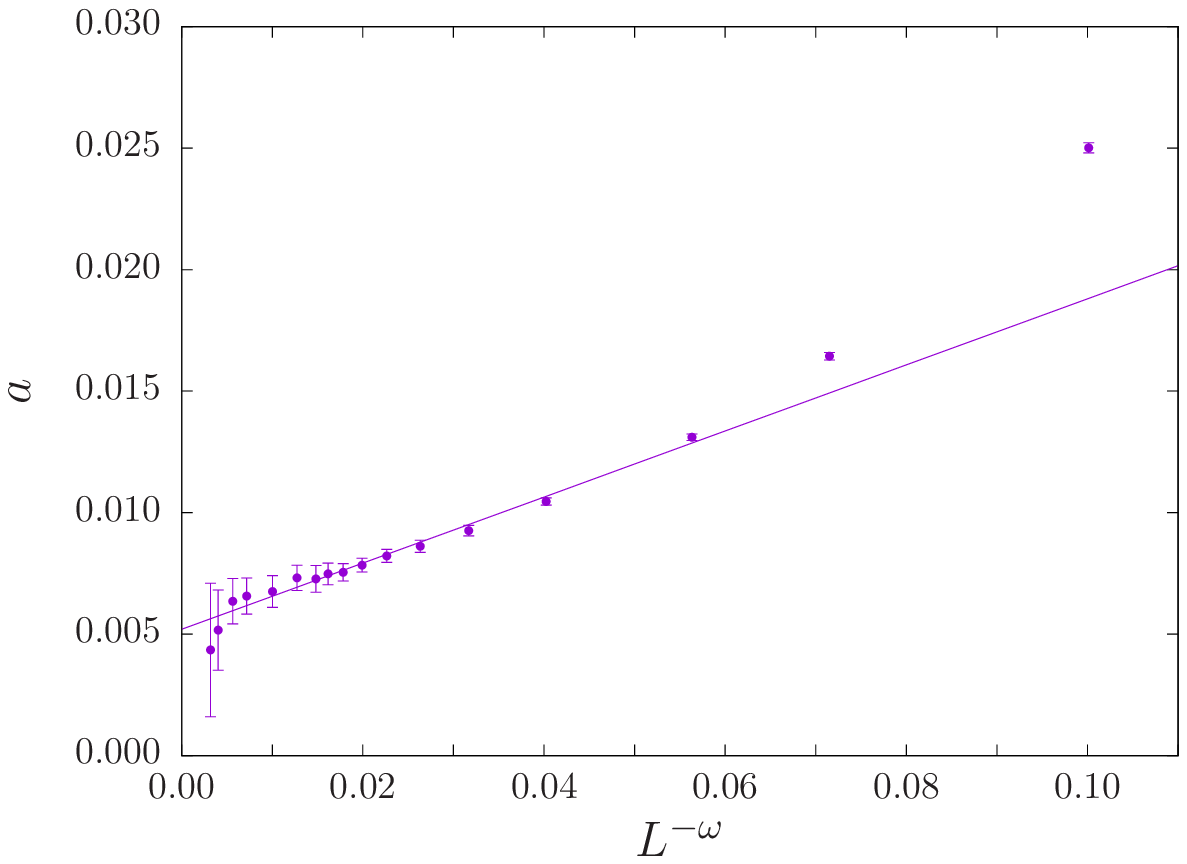}
\includegraphics [width=0.95\hsize] {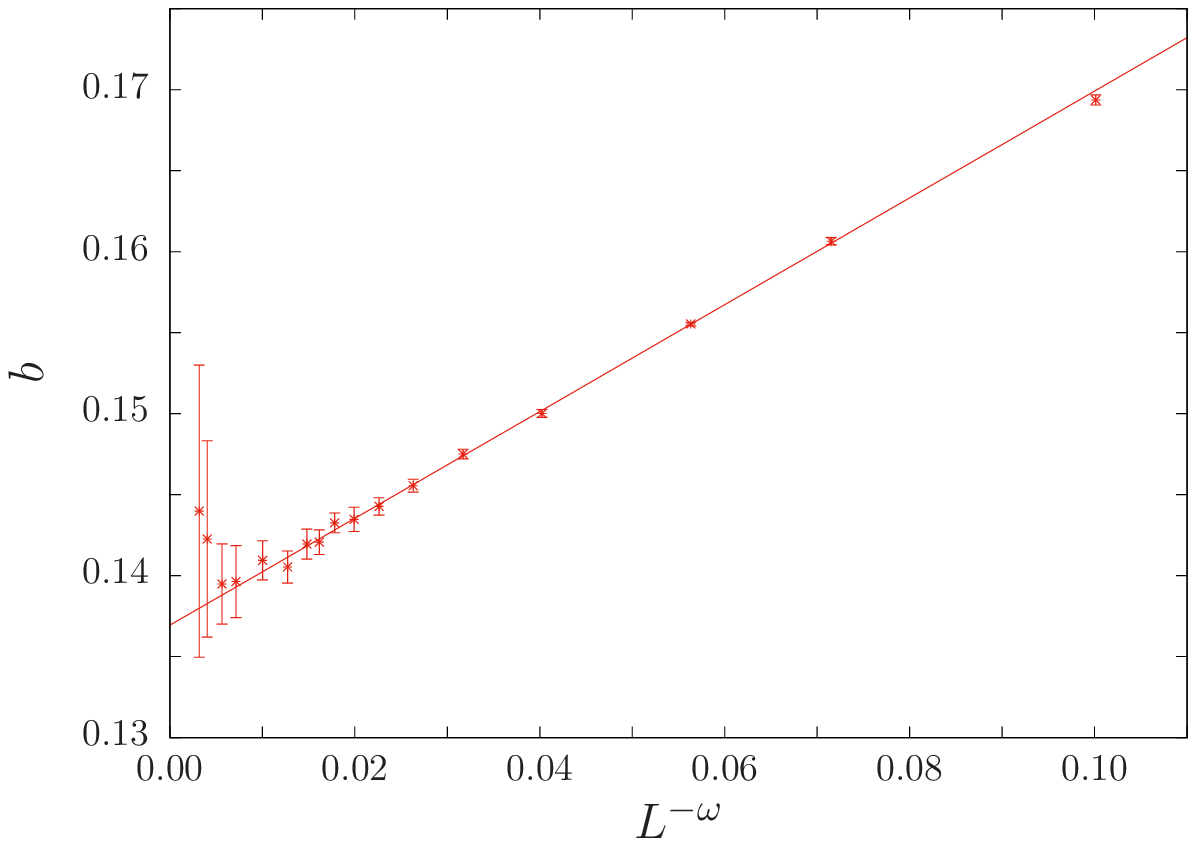}
\includegraphics [width=0.95\hsize] {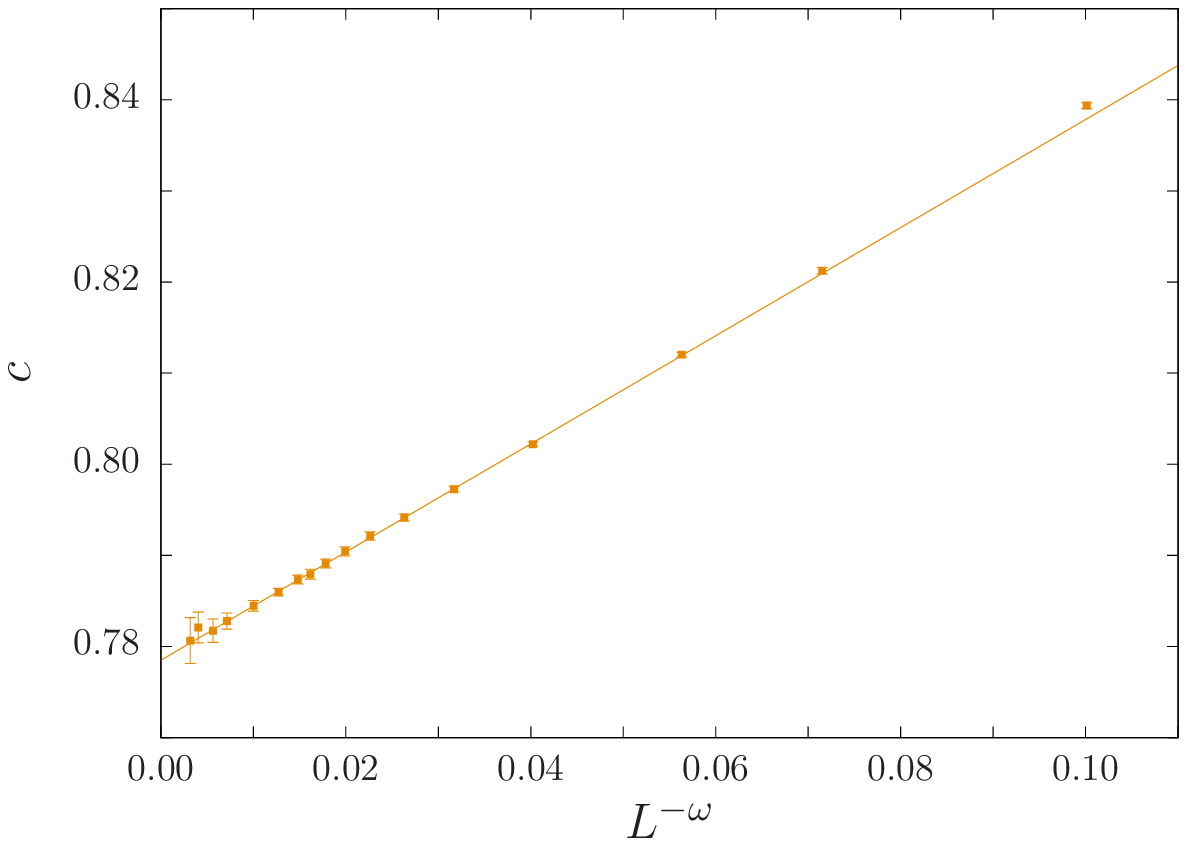}
\caption{Variation of the fit parameters $a$, $b$, and $c$ for ansatz
  Eq.~(\ref{ansatz2}) as functions of $L^{-\omega}$. The abscissa is
  chosen so that the leading corrections to scaling are
  linearized~\cite{Binder1981}, where $\omega =
  0.82968(23)$~\cite{Simmons-Duffin2017}. The solid lines show
  extrapolations to $L = \infty$ for $L \geq 32$.}
\label{fig_parameters_ansatz2}
\end{figure}

Fig.~\ref{fig_parameters_ansatz2} shows the results of the fit parameters
$a$, $b$, and $c$ of the probability distribution $P(m)$, approximated
by the ansatz Eq.~(\ref{ansatz2}). The horizontal axis is chosen to be
$L^{-\omega}$, where $\omega = 0.82968(23)$~\cite{Simmons-Duffin2017},
so that the leading corrections to scaling are
linearized~\cite{Binder1981}. There is an apparent deviation for $L =
768 $ and $ L = 1024$, but the error bars for those sizes are so
large that their contributions to the fit are less significant.  (There are many more ``bins'' in the histogram for very large $L$ so there are fewer entries in each bin.)  To within
statistical errors, there are noticeable finite-size effects for $a$,
$b$, and $c$. By doing extrapolations to the thermodynamic limit,
their values are estimated as follows,
\begin{equation}
\label{result2}
a = 0.0052(6), \;\; b = 0.137(1), \;\; c = 0.7786(3)
\end{equation}

Recently, a more precise estimate for $\beta / \nu = 0.5181489(10)$ was given by Ref.~\cite{Simmons-Duffin2017}. If we used this more precise estimate for both ansatzs Eq.~(\ref{ansatz1}) and Eq.~(\ref{ansatz2}), we will have the same extrapolated values for the parameters.

It is now known that higher order cumulants of the magnetization can have universal values. By using the probability distribution $P(m)$ of the order parameter $m$, we can calculate ratios of moments of the magnetization that are simply related to cumulants:
\begin{align*}
Q_4 &= \langle m^4 \rangle / \langle m^2 \rangle ^2 , \\
Q_6 &= \langle m^6 \rangle / \langle m^2 \rangle ^3 , \\
Q_8 &= \langle m^8 \rangle / \langle m^2 \rangle ^4 .
\end{align*}

Of course, the estimation of the cumulants from the Monte Carlo data depends upon the entire distribution; moreover, as the order of the cumulant increases, the tails of $P(m)$ become increasingly important.  Since the tails are effectively truncated by lack of data from the simulation, small biases in cumulant estimates might arise.  For high-enough order, truncation will certainly impact the value of the cumulant,  For this reason we also generated some large lattice data at a slightly larger coupling and used multi-histogram reweighing to obtain an improved estimate of the contribution of the wings.

Results are shown in Table~\ref{table_Q4_Q6}. Eqs.~(\ref{ansatz1}, \ref{result1}) and Eqs.~(\ref{ansatz2}, \ref{result2}) are used together. Error bars are estimated by using the propagation of uncertainty with correlation included (covariances between parameters $a$, $b$, and $c$ are taken into account).

As can be seen in Table~\ref{table_corrected_Q}, we find very small, systematic shifts in the estimates for $Q_4$ and $Q_6$ that are within the respective error bars of the corrected and uncorrected values.    For $Q_8$, however, the effect of truncation exceeds the error bars by a substantial amount.  Clearly, the estimation of high order moment ratios is not possible without substantially better statistics in the wings

The estimates for $Q_4$ and $Q_6$ by Eqs.~(\ref{ansatz1}, \ref{result1}) are consistent with those from the extrapolations of our MC data, Ref.~\cite{pm_Tsypin} and Ref.~\cite{hasenbusch2010}. Although the estimates by Eqs.~(\ref{ansatz2}, \ref{result2}) are higher than those from our MC data and Ref.~\cite{hasenbusch2010}, they still agree with each other to within two error bars. This might be because the estimates of $a$ and $b$ bend off for large system sizes, but the error bars for those sizes are so large that it is not possible to draw a further conclusion.

\begin{table*} []
\caption {Results for $Q_4$ and $Q_6$.}
\label{table_Q4_Q6}
\begin{center}
\begin{tabular}{@{\hspace{1em}} c @{\hspace{2em}} c @{\hspace{2em}} c @{\hspace{1em}}}
\hline
\hline
 & $Q_4$ & $Q_6$ \\
\hline
Eqs.~(\ref{ansatz1}, \ref{result1})  & 1.603 60 (13) & 3.105 55 (62) \\
Eqs.~(\ref{ansatz2}, \ref{result2})  & 1.603 97 (21) & 3.107 4 (12) \\
MC data & 1.603 52 (14) & 3.105 19 (62) \\
Typsin and Bl\"ote (2000)~\cite{pm_Tsypin}  & 1.603 99 (66) & 3.106 7 (30) \\
Hasenbusch (2010)~\cite{hasenbusch2010}  & 1.603 6 (1) & 3.105 3 (5) \\
\hline
\hline
\end{tabular}
\end{center}
\end{table*}

\begin{table*} []
\caption {Corrected and uncorrected estimates for $Q_4$, $Q_6$ and $Q_8$ at $L = 512, 768$ and 1024.}
\label{table_corrected_Q}
\begin{center}
\begin{tabular}{@{\hspace{0em}} c @{\hspace{1em}} c @{\hspace{0.7em}} c @{\hspace{0.7em}} c @{\hspace{1em}} c @{\hspace{0.7em}} c @{\hspace{0.7em}} c @{\hspace{1em}} c @{\hspace{0.7em}} c @{\hspace{0.7em}} c @{\hspace{0em}}}
\hline
\hline
& & $Q_4$ & & & $Q_6$ & & & $Q_8$ & \\
$L$ & Corrected & Uncorrected & $t$ Statistic & Corrected & Uncorrected & $t$ Statistic & Corrected & Uncorrected & $t$ Statistic \\
\hline
512	& 1.602 28(10) &	1.602 18(10) &	-0.707 1 &	3.099 10(46) &	3.098 69(45) &	-0.637 1 &	6.757 2(17) &	6.754 4(16) &	-1.199 4 \\
768	& 1.602 22(15) &	1.601 98(15) &	-1.131 4 &	3.098 97(68) &	3.098 14(68) &	-0.863 1 &	6.758 2(24) &	6.753 4(24) &	-1.414 2 \\
1024 & 1.601 97(23) & 1.601 49(22) &	-1.508 1 &	3.098 1(11) &	3.096 2(10) &	-1.278 1 &	6.769 0(37) &	6.746 8(36) &	-4.300 4 \\

\hline
\hline
\end{tabular}
\end{center}
\end{table*}

Comparing the results of fitting to the two ansatzes,
Eq.~(\ref{ansatz1}) and Eq.~(\ref{ansatz2}), one can see that the
estimates for $c$ from both fits agree with each other to within error
bars. However, the value of $b$ determined for Eq.~(\ref{ansatz1}) is
larger than that for Eq.~(\ref{ansatz2}). We believe that this is a
consequence of the correction term corresponding to $b$ in
Eq.~(\ref{ansatz1}) attempting to account for additional finite-size
corrections which are addressed explicitly by the term corresponding
to $a$ in Eq.~(\ref{ansatz2}).

\begin{figure}[]
\centering
\includegraphics [width=0.95\hsize] {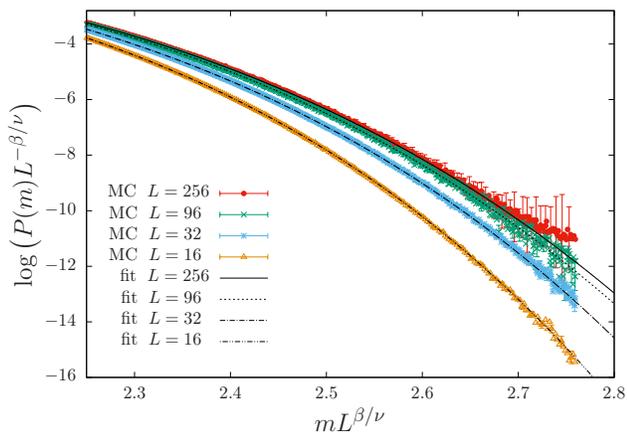}
\caption{(color online) Logarithm of the tail of the probability distribution of the order parameter (average of the left and right tails), where $m L^{\beta / \nu} \geq 2.25$, for different lattice sizes $L$. The lines are the best fits to Eq.~(\ref{ansatz2}). Curves from the top to the bottom correspond to lattice sizes $L = 256$, $96$, $32$, and $16$ respectively.}
\label{fig_pm_tail}
\end{figure}

In addition, Fig.~\ref{fig_pm_tail} shows the logarithm of the tail of
the order parameter probability distribution, where $mL^{\beta / \nu} \geq 2.25$. The values of the MC data are the averages
of the left and right tails. The solid lines are the best fits to
Eq.~(\ref{ansatz2}). The tail data for $L = 256$ fluctuate too much
to present clearly in the figure. Therefore, we applied a smoothing technique,
where each data point is the mid-point of a linear fit to 10 sequential points.  
The shape of the scaled probability distribution differs
noticeably from the thermodynamic limit, as there are non-negligible
corrections to scaling. The values of $P(m)$ are small in the tail
region, and their statistical errors are relatively high, thus, data
in the tails contribute less to the fit than those near the peaks. Although their contributions
are less significant, Fig.~\ref{fig_pm_tail} still indicates that the
fit by Eq.~(\ref{ansatz2}) performs relatively well in the tail region, at least for $mL^{\beta / \nu} \leq 2.75$. 

Overall, we have observed that the functional form
Eq.~(\ref{ansatz2}) permits a high quality, nonlinear least-squares fit
to the $P(m)$ data. Although the quality of fit for
Eq.~(\ref{ansatz1}) is reasonable for large lattice sizes, it is poor
for small lattice sizes.  The addition of a correction term
(Eq.~(\ref{ansatz2})) allows for a high-quality fit for $P(m)$ over a larger
range of system sizes.  We have observed a noticeable finite-size effect
for the fit parameters $a$, $b$, and $c$, thus Eq.~(\ref{ansatz2}) is
a high-resolution approximation expression for $P(m)$ in the
thermodynamic limit.

\vspace{2em}
\section{\label{sec:level4}conclusion}

We have determined the probability distribution $P(m)$ of the
order parameter $m$ for the simple cubic Ising model with periodic
boundary conditions at the critical temperature in a high-resolution
manner. The high quality of the distribution permitted us to obtain 
a precise functional form to describe $P(m)$ in the thermodynamic
limit as given by Eq.~({\ref{ansatz2}}). The universal
parameters of Eq.~(\ref{ansatz2}) have been determined as $a =
0.0052(6)$, $b = 0.137(1)$, and $c = 0.7786(3)$. This expression for $P(m)$
and its parameters provide a valuable benchmark for comparison with results for other models presumed to be in the Ising universality class.

\begin{acknowledgments}

We thank Dr. S.-H. Tsai for valuable discussions. Computing resources
were provided by the Georgia Advanced Computing Resource Center, the
Ohio Supercomputing Center, and the Miami University Computer Center. 

\end{acknowledgments}


\bibliography{pm}

\end{document}